# Evidence-based Hand Hygiene – Can You Trust the Fluorescent-based Assessment Methods?

**Száva Bánsághi[1,2], Viola Sári[3], Péter Szerémy[4], Ákos Lehotsky[5] Bence Takács[2], Brigitta K. Tóth[6] and Tamás Haidegger[2,7]***

[1] Semmelweis University, Doctoral School,
Üllői út 26, H-1085 Budapest, Hungary,
bansaghi.szava@phd.semmelweis.hu

[2] Óbuda University, University Research and Innovation Center (EKIK),
Bécsi út 96/b, H-1034 Budapest, Hungary,
szava.bansaghi@irob.uni-obuda.hu

[3] Budapest University of Technology and Economics, Faculty of Chemical Technology and Biotechnology,
Műegyetem rakpart 3, H-1111 Budapest, Hungary,
viola.sari@handinscan.com

[4] University of Szeged, Faculty of Medicine,
Dugonics tér 13, H-6720 Szeged, Hungary,
szeremy.peter.gabor@stud.u-szeged.hu

[5] National Institute of Oncology,
Ráth György utca 7-9, H-1122 Budapest, Hungary,
lehotsky@ooi.hu

[6] Budapest University of Technology and Economics, Faculty of Civil Engineering,
Műegyetem rakpart 3, H-1111 Budapest, Hungary,
toth.brigitta@emk.bme.hu

[7] Austrian Center for Medical Innovation and Technology (ACMIT),
Viktor Kaplan-Straße 2, 2700 Wiener Neustadt, Austria,
tamas.haidegger@acmit.at

* Correspondence: haidegger@irob.uni-obuda.hu

*Abstract: Healthcare-Associated Infections (HAI) present a major threat to patient safety globally. According to studies, more than 50% of HAI could be prevented by proper hand hygiene. Effectiveness of hand hygiene is regularly evaluated with the fluorescent method: performing hand hygiene with a handrub containing an ultra violet (UV) fluorescent marker. Typically, human experts evaluate the hands under UV-A light, and decide whether the applied handrub covered the whole hand surface. The aim of this study was to investigate*

*how different experts judge the same UV-pattern, and compare that to microbiology for objective validation. Hands of volunteer participants were contaminated with high concentration of a Staphylococcus epidermidis suspension. Hands were incompletely disinfected with UV-labeled handrub. Four different UV-box type devices were used to take CCD pictures of the hands under UV light. Next, hands were pressed to a custom-developed, hand-size agar plate for microbiological sampling. Size of inadequately disinfected areas on the hands were determined in two different ways. First, based on microbiology; the areas where colonies were grown were measured. Second, four independent senior infection control specialists were asked to mark the "missed" areas on printed image, captured under UV light. 8 hands of healthy volunteers were examined. Expert evaluations were highly uncorrelated (regarding interrater reliability) and inconsistent. The biggest difference between the human assessments was found in the case of hand #6, where properly covered hand surface annotation spanned from 21.6% to 61.1%. Microbiology results weakly correlated with the expert evaluations. In half of the cases, there were more than 10% difference in the size of properly disinfected area, as measured by microbiology versus human experts. Considering the result of the expert evaluations, variability was disconcertingly high. Evaluating the fluorescent method is challenging, even for highly experienced professionals. A patient safety quality assurance system cannot be built on these data quality. Digital tools, software-based assessment, microbiology cultivation and other objective methods should be employed.*

# 1 Introduction

Ignaz Semmelweis ordered hand washing with chlorine lime for the healthcare workers at his neonatal department in 1847, and with that he dramatically decreased the infections rates and mortality caused by childbed fever [1]. 175 years passed, but healthcare-associated infections (HAI) are still a major threats to patient safety. On any given day, 1 in 30 hospital patients acquire a HAI [2], which not only causes unnecessary suffering to patients, but also means a significant cost to hospitals. According to a meta-analysis in 2012, a surgical site infection costs 20 000 USD, while a ventilator-associated pneumonia costs 40 000 USD [3]. At least half of HAIs are preventable, and the most effective, simplest and cheapest way of intervention is proper hand hygiene [4].

Hand hygiene can only be effective if the whole surface of the hand is covered by a highly efficient (e.g., alcohol-based) handrub. 40 years ago, Taylor proved that in-clinic hand hygiene technique is often poor, the thumbs and the fingertips are the most frequently missed areas [5]. There were almost no improvements since the most frequently missed areas are still the thumbs and the fingertips according to the latest studies [6-10].

The fluorescent method has been used for over thirty years for training and monitoring hand hygiene technique in healthcare practice [11], and it is still a popular and widespread method [12-14]. First, hands should be disinfected with a handrub that contains a fluorescent additive. That additive is not visible under normal light, but it glows when hands are placed under (harmless) UV-A. Hand surface areas missed during hand hygiene – where the handrub was not rubbed – remain dark.

To assist the fluorescent method, the so-called "black boxes" were designed to exclude normal light, as evaluation is easier in dark. UV-boxes usually use a UV lamp (around 368 nm wavelength). Some devices have a fixed viewpoint, or pictures can be recorded with a camera. In their everyday educational use, there is a need for a human expert, who evaluates the hand hygiene technique and provides feedback to the staff [15]. In most cases, the results are not documented. There is a growing demand for a rapid and reproducible way to evaluate the effectiveness of hand hygiene technique [16], especially since the outbreak of the coronavirus pandemic [17]. This is also in line with the recommendations of the novel ISO 23447 international standard on hand hygiene performance and compliance [18].

The aim of this research was to test the objectivity of the human expert-based evaluation of the UV-method; to investigate the reliability of the fluorescent method. This is a major challenge since the proper hand hygiene method has a key role in the fight against healthcare-associated infections.

## 2　Materials and Methods

In this study, we compared the results of human expert-based evaluation and microbiology. We artificially contaminated hands with high concentration bacterial suspension, and then partially disinfected them with UV-marked disinfectant. Hands were sampled microbiologically, and at the same time images were recorded under UV-A light. These images were evaluated by infection control experts. Results of the different methods were compared analytically.

### 2.1　Hand Contamination

Hands of volunteer participants were disinfected with alcohol-based handrub (Sterillium, BODE-Hartmann), as during surgical hand preparation (5x1 minute). After the handrub were completely dried, hands were artificially contaminated; 1 ml 0.5 McF *Staphylococcus epidermidis* (ATCC12228) suspension were pipetted to the hands, and were evenly distributed on the whole hand surface by the participants. *Staphylococcus epidermidis* was chosen as it is part of the normal human hand flora. It is not pathogenic, hence not represented a risk for the participants [19]. The bacteria strain was provided by Biolab Zrt. (Budapest, Hungary). The initial disinfection step was necessary to completely remove the transient hand flora. Different microorganisms have different growth pattern, e.g., a yeast would easily overgrow the Staphylococcus colonies, and make the evaluation impossible. After the hands dried, they were partially disinfected. For that step, a UV-labeled handrub was used (Visirub, BODE-Hartmann was mixed in Sterillium, Bode-Chemie GmbH, Heidenheim, Germany).

## 2.2 Microbiological Sampling

Agar plates (up to 15 cm diameter) are routinely used for microbiological sampling. Taking a whole-hand size sample is challenging, as the conventional agar plates are rigid, and cannot follow the surface of the hand. We used a special sampling agar plate that was specially designed and manufactured directly for this experiment. A memory-foam was placed to a hand-shaped 3D printed frame, and was poured with a nonselective culture medium. The composition of the culture media was the following:

- Nutrient substrate 20.0 g/l
- Carbohydrates 5.0 g/l
- Macro elements 5.0 g/l
- Growth factors 5.0 g/l
- Buffers 3.0 g/l
- Triphenyl-tetrazolium-chloride 0.1 g/l
- Agar 19.9 g/l

Partially contaminated hands were sampled by the special, hand-shape agar plate. The special plate was suitable to sample the whole hand surface; the memory-foam ensured that the entire surface of the hand can be pressed to the agar without breaking it. After sampling, plates were cultured at 37°C for 24 hours, which is the standard method for culturing *Staphylococcus epidermidis*. Red colonies were grown, where hands were not appropriately decontaminated, while no colonies were formed where hand surface was disinfected. After 24-hour incubation, the colonies from the non-disinfected areas grow to a contiguous (red) area. The quality of the hand disinfection was judged based on the formed colonies.

The plates were photo-documented when the hands were pressed, and also after the incubation period, when colonies were grown (Fig. 1), clearly identifying the region of interest for evaluation. The percent of the non-properly disinfected palm surface was determined on the "After" image (Fig 1B) by software evaluation. As the contaminated areas were homogenous enough, photo editing software (paint.net, dotPDN LLC.) was used to indicate the areas having the same red color and also to give the size of these areas in pixels, considered as contaminated areas.

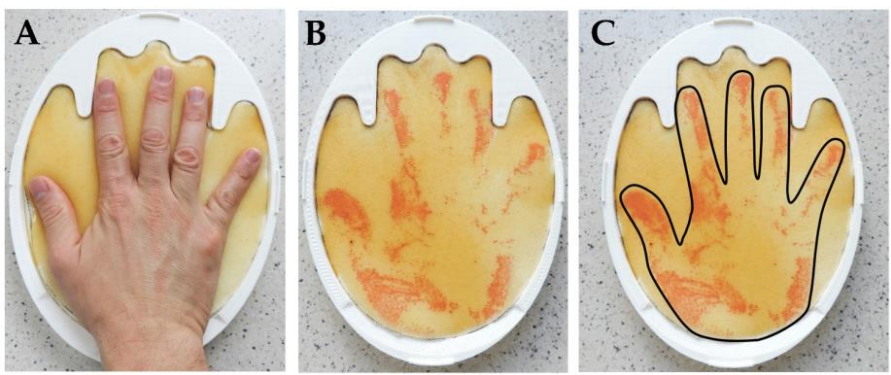

Figure 1

Microbiological sampling of the hands. Hands were artificially contaminated with *Staphylococcus epidermidis*, then partially disinfected. A) As part of the experimental method, palm was pressed to a culture media, B) Colonies grew after a 24-hour incubation period only where palm surface was not disinfected properly, C: Merged image, where the hand outline is marked (only for illustration purpose).

The hand size was calculated using the "Before" image (the picture made at the moment of pressing the palm into the agar, Fig. 1A) by the same software. Finally, the percentage of the not-properly disinfected palm area was calculated in the case of each hand.

## 2.3 UV-Boxes Supporting the Human Evaluation

Next, hands were evaluated by UV-boxes as if they were partially disinfected by a fluorescent-labeled handrub (Visirub + Sterillium, BODE-Hartmann). Comparing different UV-boxes are important, as they are different in many parameters; e.g., how well they exclude normal light, how the UV-light source is placed. These parameters may affect the quality of the evaluation.

Total of 4 boxes (Fig. 2) were employed. All four devices apply a UV-A light source, Table I summarizes their main properties. Two of the investigated UV-boxes (Derma LiteCheck Box and Schülke Optics UV Training Box) were commercially available product, routinely used in hospitals for fluorescent evaluation of hand hygiene. The Semmelweis Scanner not only records an image of the hand under UV-light but also evaluates this image by its custom own software. This feature was turned off for this study, and only the recorded, but not analyzed images were used. The Stery-Hand was a precursor of the Semmelweis Scanner. It was involved in the study because their inner design was quite different from the other three products.

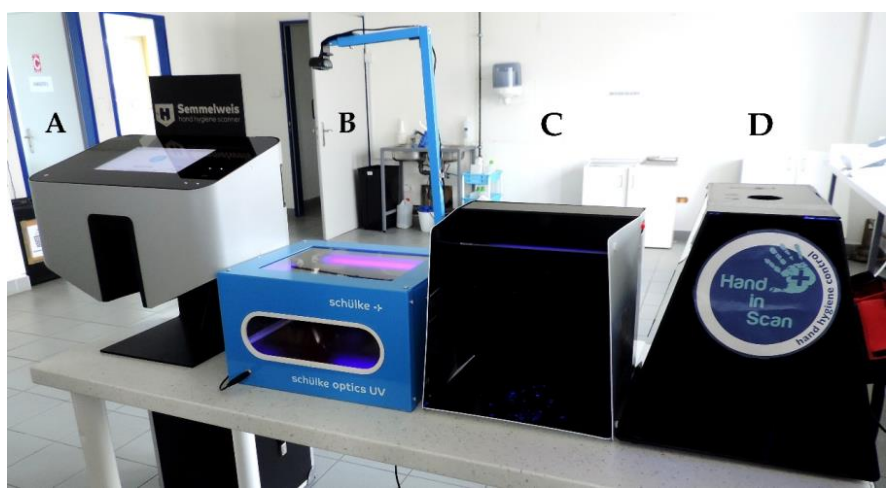

Figure 2

UV-boxes and digital health solutions compared in the study, from left to right: A) Semmelweis Scanner, B) Schülke optics UV Training Box, C) Derma LiteCheck Box, D) Stery-Hand

Table 1

Comparison of the main parameters of the 4 UV-boxes that were involved in the investigation

|  | Semmelweis Scanner | Schülke Optics UV Training Box | Derma LiteCheck Box | Stery-Hand |
|---|---|---|---|---|
| Manufacturer | HandInScan Zrt (Debrecen, Hungary) | Schülke & Mayr GmbH (Norderstedt, Germany) | Bode Chemie GmbH (Hamburg, Germany) | HandInScan Zrt (Debrecen, Hungary) |
| Dimensions (W*H*D) | 46 cm * 27 cm * 25 cm | N/A | 35 cm * 35 cm * 29.4 cm | N/A |
| Light source | 4 fluorescent LEDs | fluorescent tubes | 2 fluorescent tubes (T5 8W/BLB) | 2 fluorescent tubes |
| Wavelength | 365 nm | 366 nm | 365 nm (320 – 400 nm) | 365 nm |
| Efficacy | 4 * 1.5 W | N/A | ~ 1.2 W/m² Effective irradiance | 2 * 25 W |
| Documentation | Buil-in camera | Camera can be connected | - | Camera can be connected |

Each of the subjects' hands were placed into each of the devices, and images were recorded by CCD cameras. The Semmelweis Scanner used the built-in camera, in the case of the other three devices, the same, commercial digital camera was used (Leica D-Lux 4, Leica Camera AG, Germany). This image recording step was actually carried out before the microbiological sampling.

## 2.4　Human Experts' Evaluation

Infection control professionals were invited for evaluation, who are experts in their fields, and routinely use the fluorescent technique to monitor hand hygiene quality of healthcare-workers. They evaluated the pictures recorded by the different devices. During their everyday practice, professionals do not record the results, and only provide verbal feedback to healthcare-workers.

In this study, the recorded images were printed, laminated and provided for the expert for evaluation. Since the pictures were covered by a transparent foil, graphical evaluation was made possible. The two layers were fixed together to prevent the shifting of the foil, and have been tied to a drawing board. Experts drew around the not-sufficiently disinfected (dark) areas with a permanent marker on the foil. Each expert had only 1 minute to evaluate an image, as we assumed that the lack of time mimic better the in-clinique circumstances and thus makes the result more realistic.

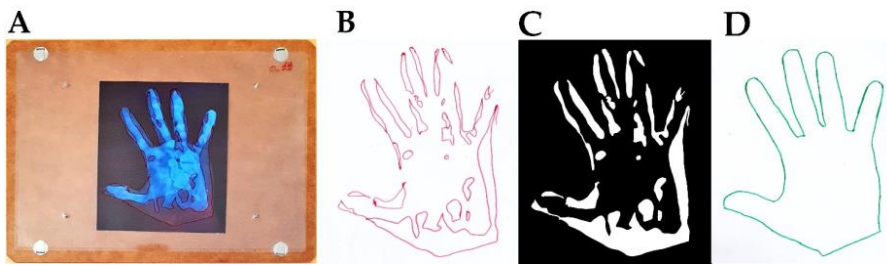

Figure 3

Expert's evaluation of the recorded images. A) Infection control professionals marked the dark, untreated areas on the image, B) Contaminated area marked by the experts, after the foil was separated from the printed image, C) Size of the marked areas in pixels was measured by the software D) hand contour was also marked to a different foil, then its area was measured in pixels

The foils marked with the evaluation acquired this way were photoscanned, and the marked (contaminated) areas were determined by software evaluation (paint.net, dotPDN LLC, WA, USA) with the built-in Flood Fill Algorithm. The palm surface was determined similarly after the hand contour was marked, using a different foil. The percentage of the insufficiently disinfected areas was determined, and compared to the same percentage calculated by the microbiological results.

## 3　Results

During the experiment, we prepared digital pictures of eight subjects, providing differently disinfected hands one by one with four different devices under UV-light.

All 8 hands were recorded by all four devices, so 32 recorded images were provided to the experts. Four infection control experts participated in the study; all four experts evaluated the 32 printed images, so finally, we had 128 evaluated pictures. The 8 hands were also sampled microbiologically.

## 3.1 Microbiological Sampling

In Figure 4, the calculation steps detailed above were presented in the example (Hand #2). On the picture made at the moment when the hand was pressed into the agar (Fig. 4A) size of the frame (3 154 773 pixels) and palm size (1 587 695 pixels) were determined. The palm covered 50.33% of the frame (1 587 695 / 3 154 773 * 100). In the next image (Fig. 4B), which was taken after the incubation period, the frame size was measured as 4 791 851 pixels. As we know the palm size: frame size ratio, in this image the palm would take 2 411 583 pixels (4 791 851 * 0.5033). On the same image, the size of the colony-covered area was measured (Fig. 4C), in this case it was 950 504 pixels. The insufficiently disinfected area was found to be 39.41% of the whole palm (950 504 / 2 411 583 * 100).

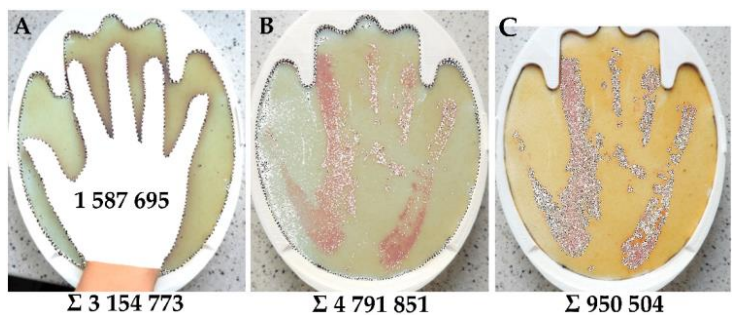

Figure 4
Sample shows how the contaminated palm areas were calculated based on the microbiological result.
A) pixel count of the hand area; B, camera image of the cultivated agar with the total area pixel count;
C, Pixel-level selection of the contaminated area

The same method was used for calculating the rate of contaminated palm area at each of the 8 samples (Table 2). This may be significantly enhanced in the future with the help of AI-based image processing [20].

Table 2
Determination of the contaminated ratio of the palm area

| Hand | Image taken during sampling | | | Image taken after incubation | | | Contaminated palm area [%] |
|---|---|---|---|---|---|---|---|
| | Palm area [pixels] | Frame area [pixels] | Hand: Frame ratio [%] | Frame area [pixels] | Calculated palm surface | Colony area [pixels] | |
| #1 | 1 879 156 | 3 533 170 | 53.19 | 5 482 974 | 2 916 181 | 913 870 | 31.34 |
| #2 | 1 587 695 | 3 154 773 | 50.33 | 4 791 851 | 2 411 583 | 950 504 | 39.41 |

| #3 | 3 144 581 | 5 823 635 | 54.00 | 5 320 196 | 2 872 740 | 659 192 | 22.95 |
| #4 | 4 236 541 | 8 764 899 | 48.34 | 4 824 464 | 2 331 920 | 759 125 | 32.55 |
| #5 | 2 794 126 | 4 413 859 | 63.30 | 4 471 771 | 2 830 786 | 733 862 | 25.92 |
| #6 | 2 296 875 | 3 740 697 | 61.40 | 4 845 982 | 2 975 546 | 676 739 | 22.74 |
| #7 | 2 888 095 | 4 586 772 | 62.97 | 5 068 097 | 3 191 165 | 1 417 681 | 44.43 |
| #8 | 2 561 868 | 4 197 213 | 61.04 | 4 990 252 | 3 045 918 | 969 693 | 31.84 |

### 3.2 UV-Boxes

The hands that were partially disinfected by a fluorescent labeled handrub were placed to the 4 devices, under UV-A light. Figure 5 shows how differently the same hand (Hand #2) was recorded by the different devices. The main difference in these images is the intensity of the background light; well-shaded design resulted images with higher contrast.

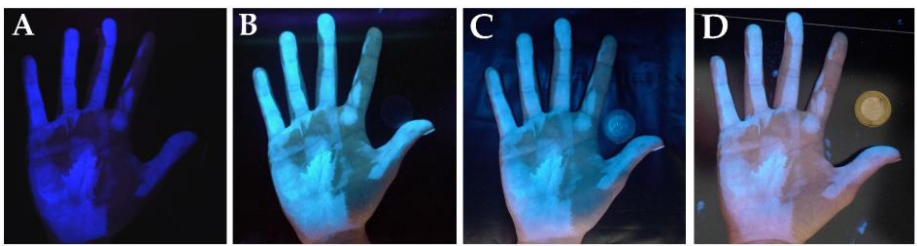

Figure 5

Same hand (Hand #2) was placed into different UV-boxes. The hand was partially treated (i.e., incompletely disinfected) with a fluorescently labeled handrub. A) Semmelweis Scanner, B) Schülke Optics UV Training Box, C) Derma LiteCheck Box, D) Stery-Hand

### 3.3 Experts' Evaluations

Experts marked the insufficiently disinfected palm areas on the foils. After recollecting the 128 images (8 hands * 4 devices * 4 experts), the evaluator foils were photoscanned and analyzed. As Figure 6 shows, the experts marked roughly the same areas, but they differed a bit in where the border between the light and dark area was drawn. The fluorescent dye makes a color gradient on the hand, and it was not made clear to the evaluators at all which part of the hand is bright enough to declare it disinfected.

To measure the percentage of the contaminated area, the size of the marked area was measured. Figure 7 shows an example (Hand #2), how these areas were evaluated differently by experts (Fig. 7A) and by devices (Fig. 7B). Values indicate a large deviation; neither the experts nor the devices show a clear correlation with the size of the contaminated areas.

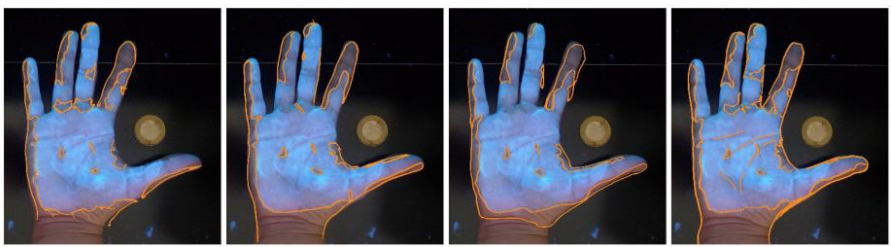

Figure 6

The same recorded images were evaluated by four infection control experts. The marked, "dark" areas were considered insufficiently disinfected.

The lowest evaluated value was 16.87% (Device #4 – Expert #3), while the highest value was more than three-times higher, 53.79% (Device #2 – Expert #3). The highest and the lowest values were taken by the same expert. Note, that all these pictures were taken from the same hand.

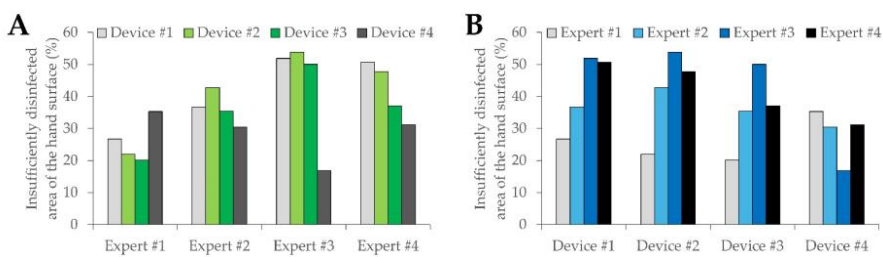

Figure 7

Same images (Hand #2) were recorded in 4 devices, and then evaluated by 4 infection control experts. The insufficiently contaminated hand surface was compared A) by experts and B) by devices.

## 3.4 Comparison of the Experts' Evaluations with the Microbiology-based Results

In Table 3, all the above calculated, decontaminated hand areas are compared; in the case of all hands, the value determined by the microbiological sampling, and also the 16 values by the expert evaluation. Figure 8 shows the same data, also the Q1, the median, and the Q3 values of expert evaluation in the case of each hand.

As it has clearly shown from the comparison, a huge deviation of the results can be seen that depends highly on the human factor. Assumed that the bacteria colonies created homogenous contamination on the whole surface of the hand, the result of the microbiologic incubation was considered the real quality of the disinfection. Note that in most cases even the median values of the expert evaluation were not even close to the result of microbiology.

Table 3
Comparison of the percentage of the not adequately disinfected areas based on the experts' evaluations and results given by the microbiologic sampling

| Hand | Device | Contamination of the palm (%) | | | | Microbiology |
| --- | --- | --- | --- | --- | --- | --- |
| | | Expert #1 | Expert #2 | Expert #3 | Expert #4 | |
| Hand #1 | Device #1 | 33.53 | 19.41 | 35.55 | 35.33 | 31.34 |
| | Device #2 | 13.60 | 24.56 | 34.31 | 36.06 | |
| | Device #3 | 19.39 | 16.44 | 32.38 | 30.45 | |
| | Device #4 | 28.56 | 12.70 | 20.85 | 30.68 | |
| Hand #2 | Device #1 | 26.67 | 36.71 | 51.89 | 50.70 | 39.41 |
| | Device #2 | 21.98 | 42.84 | 53.79 | 47.77 | |
| | Device #3 | 20.17 | 35.43 | 50.04 | 37.08 | |
| | Device #4 | 35.33 | 30.50 | 16.87 | 31.18 | |
| Hand #3 | Device #1 | 28.91 | 32.34 | 53.06 | 43.15 | 22.95 |
| | Device #2 | 34.50 | 28.86 | 46.43 | 47.66 | |
| | Device #3 | 27.21 | 26.14 | 23.57 | 43.36 | |
| | Device #4 | 26.72 | 18.41 | 32.29 | 29.32 | |
| Hand #4 | Device #1 | 32.17 | 26.18 | 36.41 | 43.43 | 32.55 |
| | Device #2 | 20.90 | 25.55 | 32.98 | 38.15 | |
| | Device #3 | 39.99 | 22.26 | 27.05 | 34.64 | |
| | Device #4 | 22.32 | 12.31 | 32.25 | 34.42 | |
| Hand #5 | Device #1 | 42.69 | 49.16 | 54.99 | 35.52 | 25.92 |
| | Device #2 | 41.81 | 32.73 | 30.30 | 40.28 | |
| | Device #3 | 55.60 | 21.38 | 43.07 | 44.01 | |
| | Device #4 | 42.56 | 24.38 | 43.05 | 29.17 | |
| Hand #6 | Device #1 | 35.69 | 25.57 | 53.35 | 42.33 | 22.74 |
| | Device #2 | 35.10 | 41.08 | 49.91 | 61.11 | |
| | Device #3 | 31.87 | 32.78 | 45.81 | 43.13 | |
| | Device #4 | 21.61 | 30.69 | 24.77 | 20.91 | |
| Hand #7 | Device #1 | 50.99 | 43.79 | 26.52 | 56.11 | 44.43 |
| | Device #2 | 51.24 | 50.58 | 32.12 | 48.65 | |
| | Device #3 | 41.82 | 38.92 | 58.48 | 56.39 | |
| | Device #4 | 39.56 | 32.91 | 39.77 | 41.72 | |
| Hand #8 | Device #1 | 40.96 | 38.10 | 49.80 | 52.09 | 31.84 |
| | Device #2 | 41.88 | 36.41 | 60.74 | 51.51 | |
| | Device #3 | 45.28 | 28.27 | 49.95 | 45.68 | |
| | Device #4 | 23.06 | 23.82 | 52.68 | 52.32 | |

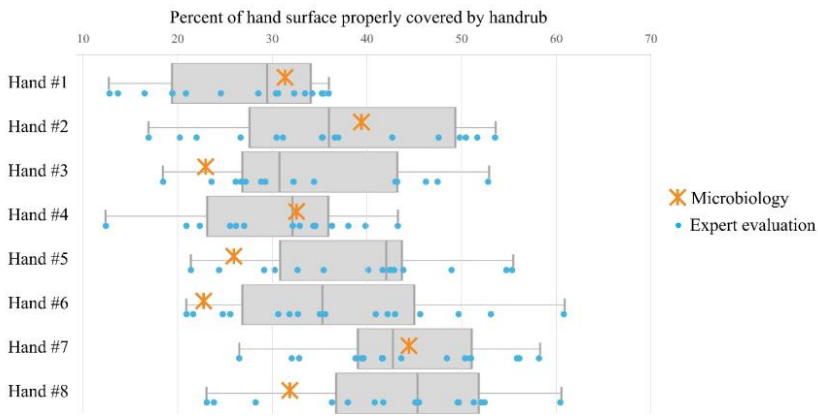

Figure 8

Percent of the insufficiently disinfected area. In the case of all investigated hand, images were recorded by 4 devices, and evaluated by four professional, so 16 different value was measured in total. Blue dots represent these individual values. The bar chart shows the Q1, the median, and the Q3 value of the 16 individual values, the error bars show the minimal and the maximal values. In the case of each hand, the contaminated hand area measured by microbiological sampling was labeled with the orange mark.

## 3 Discussion

During our experiments, we imitated the evaluations implemented as a daily routine with UV boxes. Based on our results, we can claim that the experts evaluated the given pictures with large deviation. The time pressure and the human factor carry a lot of mistake possibilities. The fluorescent method was deemed not to be an objective method. It can be used to highlight some missed areas during hand hygiene assessment on a personal level, but a hospital-wide monitoring program should not be built on such unreliable data. Automation of the entire evaluation process could be a promising alternative [21].

The main limitation of our study was that only one bacterial strain was investigated. Different germs may require different handrub concentrations. Another important limitation of the study was the limited sample size. Although we think that the huge variability in the data well supports our conclusion, a more uniformized way of data collection is required, if we aim to compare hand hygiene technique between wards or between time intervals. Nevertheless, it is foreseen that the current lasting coronavirus pandemic situation will increase the need for objective personal hygiene and assessment methods, not solely in the medical domain, but also in other professional sectors [22, 23].

## Conclusions

Even if it is managed and administered well, consistent evaluation of hand hygiene quality by the fluorescent method is not possible, as interobsever variability was found to be high, and the results were weakly correlated with microbiology-based outcomes. To build-up an evidence-based quality assurance system, hand hygiene quality should be evaluated with a more objective method, for instance, with software based assessment or microbiology cultivation.

### Acknowledgement

The authors would like to thank the Biolab Zrt., especially Dr. László Ferenci and Erzsébet Frunyó, for making possible to carry out the experiments, and for their comments and advice. Special thanks to the infection control professionals, who evaluated the recorded images, and rent their UV-boxes for the study. We also grateful to the HandInScan Zrt, for providing their products for the study. We would also like to thank the work of Dr. Ágnes Suhajda, who were the supervisor of V.S. Special thanks to Klara Haidegger for editing and formatting the manuscript. T.H. is a Bolyai fellow of the Hungarian Academy of Sciences.